\documentclass[aps,showpacs,preprintnumbers,amsmath, amssymb]{revtex4}

\usepackage{float}
\usepackage{graphics,epsfig}
\usepackage{graphicx}
\usepackage{dcolumn}
\usepackage{bm}
\usepackage{amsmath}

\begin{document}
\baselineskip=0.8 cm

\title{{\bf Stationary scalar hairy configurations supported by Neumann compact stars}}
\author{Yan Peng$^{1}$\footnote{yanpengphy@163.com}}
\affiliation{\\$^{1}$ School of Mathematical Sciences, Qufu Normal University, Qufu, Shandong 273165, China}

\vspace*{0.2cm}
\begin{abstract}
\baselineskip=0.6 cm
\begin{center}
{\bf Abstract}
\end{center}

We study stationary scalar field hairy configurations
supported by asymptotically flat horizonless compact stars.
At the star surface, we impose Neumann boundary
conditions for the scalar field.
With analytical methods, we obtain bounds
on the frequency of scalar fields.
For certain discrete frequency satisfying the bounds,
we numerically get solutions of scalar hairy stars.
We also disclose effects of model parameters on the
discrete frequency of scalar fields.

\end{abstract}

\maketitle
\newpage
\vspace*{0.2cm}

\section{Introduction}

The no hair theorem, see e.g. \cite{Bekenstein}-\cite{JBN}, plays an important role in the
development of black hole theories. The classical no hair theorem states
that asymptotically flat black holes cannot support
scalars, massive vectors and Abelian Higgs hairs outside the horizon, for recent progress
see \cite{mr1}-\cite{CW} and reviews see \cite{Bekenstein-1,CAR}.
It was believed that the no hair property is mainly
due to the existence of absorbing horizons.

Recently, it was found that such no hair behaviors
also appear in spacetimes without horizons.
Hod firstly proved that static massive scalar fields cannot exist
outside asymptotically flat horizonless neutral
compact stars, which are endowed with scalar
reflecting surface boundary conditions \cite{Hod-6}.
With static scalar fields non-minimally coupled to the gravity,
asymptotically flat horizonless neutral reflecting compact stars
still cannot support static scalar field hairs \cite{nonm1,nonm2,nonm3}.
No static scalar hair theorem also holds for horizonless
neutral reflecting compact stars with a nonzero cosmological
constant \cite{Bhattacharjee,Yan Peng-1,Yan Peng-3}.
Then what is the case when the compact star is charged.
It was found that no static scalar hair theorem holds
for large charged asymptotic flat horizonless reflecting compact stars
and in contrast, static scalar hair can form outside
small charged asymptotic flat horizonless reflecting compact stars \cite{Hod-8}-\cite{Yan Peng-8}.
Very differently, asymptotically flat charged black holes cannot support scalar hairs \cite{cbh1,cbh2}.
If we want to obtain static scalar hairs outside asymptotically flat charged black holes,
we usually have to put the black hole in a box \cite{Dolan,Basu,Oscar,Sanchis,blc1,blc2}.

From above progress, it seems that no static scalar field hair behavior
is a general property in the neutral horizonless reflecting compact stars.
Theoretically, the scalar field can be either static or stationary.
So it is also interesting to examine whether stationary scalar fields can condense
outside neutral horizonless compact stars.
In the background of black holes, stationary scalar hairs can exist
when the spacetime is rotating \cite{Hod-1}-\cite{st11}.
In contrast, for horizonless compact reflecting stars,
Hod showed that non-rotating neutral star can support exterior stationary
scalar field hairs \cite{Stationary}.
Instabilities of compact stars with Neumann boundary conditions
were also numerically investigated in \cite{Maggio}.
One naturally wonders whether this instability can trigger
scalar hairs outside Neumann compact stars.
In this work, we plan to examine whether stationary scalar fields can
condense outside neutral horizonless Neumann compact stars,
where we impose Neumann boundary conditions instead of
Dirichlet reflecting conditions in \cite{Stationary}.

In the following, we introduce the system composed of stationary scalar fields and
neutral horizonless Neumann compact stars. We analytically obtain bounds on the frequency of stationary scalar hairs.
We also numerically get stationary hairy neutral horizonless Neumann
compact stars. We summarize main results in the last section.

\section{Bounds on the frequency of stationary scalar fields}

We consider the possible configuration with exterior
stationary scalar fields outside compact stars.
And the corresponding matter field Lagrange density is given by
\begin{eqnarray}\label{lagrange-1}
\mathcal{L}=-|\nabla_{\mu} \Psi|^{2}-m^{2}\Psi^{2}.
\end{eqnarray}
Here $\Psi$ is the stationary scalar field and m
corresponds to the scalar field mass.

In Schwarzschild coordinates, the spacetime
outside the compact star is described by \cite{metric}
\begin{eqnarray}\label{AdSBH}
ds^{2}&=&-f(r)dt^{2}+\frac{dr^{2}}{f(r)}+r^{2}(d\theta^{2}+sin^{2}\theta d\phi^{2}),
\end{eqnarray}
where $f(r)$ is the metric function.
We take the radial coordinate $r=r_{s}$ as the radius of the star surface.
Outside the surface, it is the Schwarzschild type
solution $f(r)=1-\frac{2M}{r}$, where M is the star mass.
Since we focus on horizonless star, there is the relation $r_{s}>2M$.
$\theta$ and $\phi$ are angular coordinates.

The scalar field equation is \cite{ISP,ISPY,metric3}
\begin{eqnarray}\label{BHg}
(\nabla^{\nu}\nabla_{\nu}-m^2)\Psi=0.
\end{eqnarray}

We study stationary scalar fields expressed in the simple form
\begin{eqnarray}\label{BHg}
\Psi(t,r)=e^{-i \omega t}\psi(r)
\end{eqnarray}
with $\omega$ as the frequency.

Then the equation (3) can be expressed as
\begin{eqnarray}\label{BHg}
\psi''+(\frac{2}{r}+\frac{f'}{f})\psi'+(\frac{\omega^2}{f^2}-\frac{m^2}{f})\psi=0
\end{eqnarray}
with $f=1-\frac{2M}{r}$.

At the surface, we impose the Neumann condition for
the scalar field in the form $\psi'(r_{s})=0$.
At the infinity, the general asymptotic behavior for
the bounded scalar field is \cite{BC}
\begin{eqnarray}\label{BHg}
\psi\sim  \frac{1}{r}e^{-\sqrt{m^2-\omega^2}r}.
\end{eqnarray}

So the scalar field satisfies boundary conditions
\begin{eqnarray}\label{InfBH}
&&\psi'(r_{s})=0,~~~~~~~~~~~~~\psi(\infty)=0.
\end{eqnarray}

With a new radial function $\tilde{\psi}=\sqrt{r}\psi$,
the scalar field equation (5) can be transformed into
\begin{eqnarray}\label{BHg}
r^2\tilde{\psi}''+(r+\frac{r^2f'}{f})\tilde{\psi}'+(-\frac{1}{4}-\frac{rf'}{2f}+\frac{\omega^2r^2}{f^2}-\frac{m^2r^2}{f})\tilde{\psi}=0
\end{eqnarray}
with $f=1-\frac{2M}{r}$.

Boundary conditions of the function $\tilde{\psi}$ are
\begin{eqnarray}\label{InfBH}
&&\tilde{\psi}'(r_{s})=\frac{1}{2\sqrt{r_{s}}}\psi(r_{s})+\sqrt{r_{s}}\psi'(r_{s})=\frac{1}{2\sqrt{r_{s}}}\psi(r_{s})
=\frac{1}{2r_{s}}\tilde{\psi}(r_{s}),
\end{eqnarray}
\begin{eqnarray}\label{InfBH}
\tilde{\psi}\sim \frac{1}{\sqrt{r}}e^{-\sqrt{m^2-\omega^2}r}\rightarrow 0~~~as~~~r\rightarrow \infty.
\end{eqnarray}

In the following, we divide our discussion into three cases:
$\tilde{\psi}(r_{s})=0$, $\tilde{\psi}(r_{s})>0$ and $\tilde{\psi}(r_{s})<0$.

In the first case of $\tilde{\psi}(r_{s})=0$, considering the relation (10),
at least one extremum point $r_{peak}$ of $\tilde{\psi}(r)$ exists
between the surface and the infinity boundary.
The extremum point may be positive maximum extremum point satisfying
\begin{eqnarray}\label{InfBH}
\tilde{\psi}(r_{peak})>0, \tilde{\psi}'(r_{peak})=0, \tilde{\psi}''(r_{peak})\leqslant0
\end{eqnarray}
or negative minimum extremum point with
\begin{eqnarray}\label{InfBH}
\tilde{\psi}(r_{peak})<0, \tilde{\psi}'(r_{peak})=0, \tilde{\psi}''(r_{peak})\geqslant0.
\end{eqnarray}

In the second case of $\tilde{\psi}(r_{s})>0$, there is $\tilde{\psi}'(r_{s})=\frac{1}{2r_{s}}\tilde{\psi}(r_{s})>0$.
Around the star surface, the positive scalar field firstly increases to be larger and
approaches zero at the infinity. So one positive maximum extremum
point $r_{peak}$ exists. At this maximum extremum point, we have the relation
\begin{eqnarray}\label{InfBH}
\tilde{\psi}(r_{peak})>0, \tilde{\psi}'(r_{peak})=0, \tilde{\psi}''(r_{peak})\leqslant0.
\end{eqnarray}

In the third case of $\tilde{\psi}(r_{s})<0$, there is $\tilde{\psi}'(r_{s})=\frac{1}{2r_{s}}\tilde{\psi}(r_{s})<0$.
Around the star surface, the negative scalar field firstly decreases to be more negative and
approaches zero at the infinity. So one negative minimum extremum
point $r_{peak}$ exists. At this minimum extremum point, we have the relation
\begin{eqnarray}\label{InfBH}
\tilde{\psi}(r_{peak})<0, \tilde{\psi}'(r_{peak})=0, \tilde{\psi}''(r_{peak})\geqslant0.
\end{eqnarray}

As a summary of (11-14), scalar field solutions satisfy characteristic relations
\begin{eqnarray}\label{InfBH}
\{ \tilde{\psi}'=0~~~~and~~~~\tilde{\psi} \tilde{\psi}''\leqslant0\}~~~~for~~~~r=r_{peak}.
\end{eqnarray}

From (8), we get the equation
\begin{eqnarray}\label{BHg}
r^2\tilde{\psi}\tilde{\psi}''+(r+\frac{r^2f'}{f})\tilde{\psi}\tilde{\psi}'+(-\frac{1}{4}-
\frac{rf'}{2f}+\frac{\omega^2r^2}{f^2}-\frac{m^2r^2}{f})\tilde{\psi}^{2}=0.
\end{eqnarray}

Relations (15) and (16) lead to the inequality
\begin{eqnarray}\label{BHg}
-\frac{1}{4}-\frac{rf'}{2f}+\frac{\omega^2r^2}{f^2}-\frac{m^2r^2}{f}\geqslant0~~~for~~~r=r_{peak}.
\end{eqnarray}

With $f(r)=1-\frac{2M}{r}$, (17) can be transformed into
\begin{eqnarray}\label{BHg}
\frac{1}{r^2}-\frac{8m^2M}{r}+4(m^2-\omega^2)\leqslant0~~~for~~~r=r_{peak}.
\end{eqnarray}

(18) yields the relation
\begin{eqnarray}\label{BHg}
(\frac{1}{r}-4m^2M)^2+4(m^2-\omega^2)-16m^4M^2\leqslant0~~~for~~~r=r_{peak}.
\end{eqnarray}

From (19), we obtain the inequality
\begin{eqnarray}\label{BHg}
4(m^2-\omega^2)-16m^4M^2\leqslant 0.
\end{eqnarray}

It yields the bound for the frequency as
\begin{eqnarray}\label{BHg}
\frac{\omega^2}{m^2}\geqslant 1-4m^2M^2.
\end{eqnarray}

According to (18), there is
\begin{eqnarray}\label{BHg}
-\frac{8m^2M}{r}+4(m^2-\omega^2)\leqslant0~~~for~~~r=r_{peak}.
\end{eqnarray}

From (22), we get the relation
\begin{eqnarray}\label{BHg}
\frac{\omega^2}{m^2}\geqslant 1-\frac{2M}{r_{peak}}\geqslant 1-\frac{2M}{r_{s}}.
\end{eqnarray}

According to the infinity asymptotic behavior (6),
the bound-state fields are characterized by proper
resonant frequencies with the relation \cite{Hod-9}
\begin{eqnarray}\label{BHg}
m^{2}\geqslant \omega^{2}.
\end{eqnarray}

With (21), (23) and (24), we obtain bounds on the frequency as
\begin{eqnarray}\label{BHg}
max \{1-\frac{2M}{r_{s}},1-4m^2M^2\} \leqslant\frac{\omega^2}{m^2}\leqslant 1.
\end{eqnarray}

In the limit of $r_{s}\gg M$, for fixed nonzero $mM$, (25) is the
same as Hod's bound \cite{Stationary}
\begin{eqnarray}\label{BHg}
1-\frac{2M}{r_{s}} \leqslant\frac{\omega^2}{m^2}\leqslant 1.
\end{eqnarray}

For ultra-compact star, $r_{s}$ can be chosen a little larger than 2M.
In the limit $r_{s}\rightarrow 2M$, (25) is equal to
\begin{eqnarray}\label{BHg}
1-4m^2M^2 \leqslant\frac{\omega^2}{m^2}\leqslant 1.
\end{eqnarray}

\section{Stationary scalar configurations supported by neutral Neumann compact stars}

The stationary scalar hairy compact stars were analytically studied in the limit of $M\ll r_{s}$,
where the Dirichlet reflecting conditions were imposed at the star surface \cite{Stationary}.
With numerical methods, we carry out the discussion
in the case of relaxing the condition $M\ll r_{s}$.
In this work, we try to search for stationary scalar hairy compact stars
with Neumann surface boundary conditions.

Since the equation (5) is of the second order, we need values
$\psi(r_{s})$, $\psi'(r_{s})$ and model parameters to integrate the equation.
We choose to describe the system with dimensionless quantity $mr_{s}$,~$mM$ and $\frac{\omega}{m}$
according to the symmetry of the equation (5) in the form
\begin{eqnarray}\label{BHg}
r\rightarrow k r,~~~~ m\rightarrow m/k,~~~~ M\rightarrow k M,~~~~\omega\rightarrow \omega/k.
\end{eqnarray}
Neumann surface boundary conditions yield $\psi'(r_{s})=0$.
Using the symmetry $\psi\rightarrow k \psi$ of the equation (5), we can fix $\psi(r_{s})=1$.
For fixed $mr_{s}$ and $mM$, we can search for the proper $\frac{\omega}{m}$
satisfying the infinity boundary condition $\psi(\infty)=0$.

With given values of $mr_{s}$ and $mM$, we obtain discrete $\frac{\omega}{m}$
satisfying the infinity boundary condition.
For example, for $mr_{s}=5$ and $mM=1.5$, we find $\frac{\omega}{m}$ is around 0.91654092 satisfying the bounds (25).
Now we plot the scalar field solution with $mr_{s}=5$,~$mM=1.5$ and $\frac{\omega}{m}=0.91654092$ in the left panel of Fig. 1.
As shown in the picture, the scalar field satisfying Neumann
conditions at the star surface and asymptotically approaches zero at the infinity.
In addition, for $mr_{s}=5$, $mM=1.5$ and $\frac{\omega}{m}=0.96407075$, we obtain configurations with nodes
and plotted it in the right panel of Fig. 1. In fact, we can also get configurations
with many nodes. In black hole backgrounds, configurations without nodes
are usually thermodynamically more stable compared to
configurations with nodes \cite{st1,st2}. In this work, we focus on configurations without nodes.
We will try to examine stability of such horizonless
configurations with nodes in the next work.

\begin{figure}[h]
\includegraphics[width=200pt]{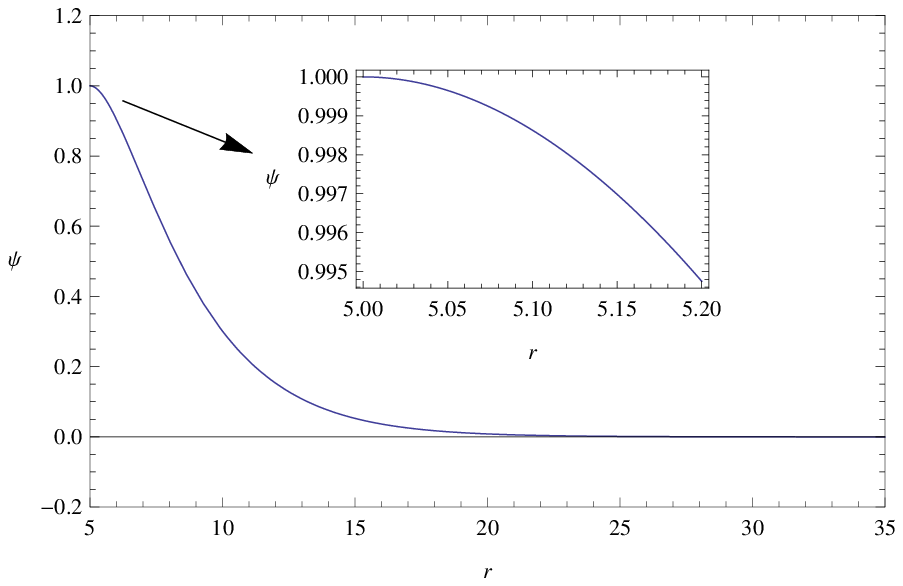}\
\includegraphics[width=200pt]{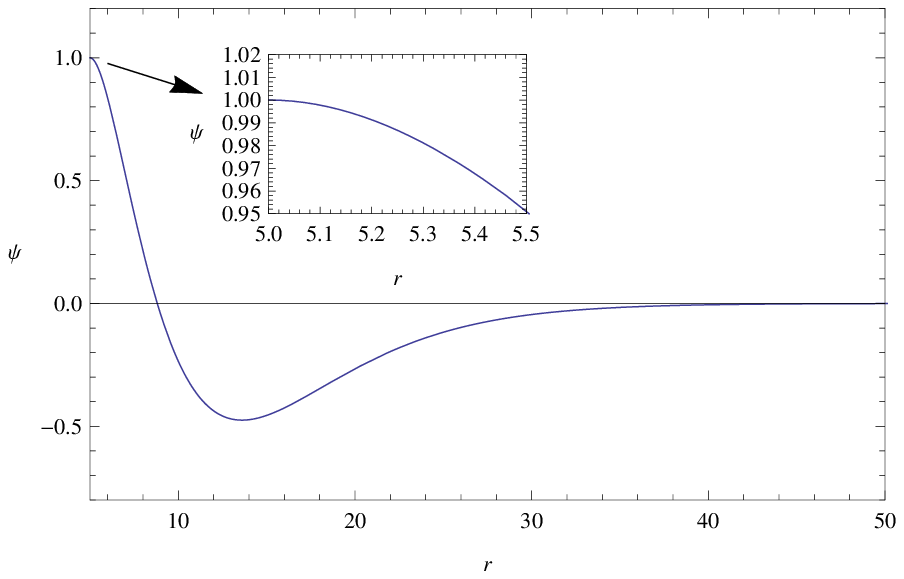}\
\caption{\label{EEntropySoliton} (Color online) We plot the scalar field function $\psi$
with respect to the radial coordinate $r$ in the case of $mr_{s}=5$ and $mM=1.5$. The left panel is with $\frac{\omega}{m}=0.91654092$
and the right panel corresponds to $\frac{\omega}{m}=0.96407075$.}
\end{figure}

In the limit of $r_{s}\gg M$, according to
the bounds (26), the numerical value of the frequency should be almost 1.
In this work, since we relax the condition $r_{s}\gg M$, the numerical value of the
frequency can be away from 1.
In the following, we study the frequency of the stationary scalar hair.
In Table I, we show cases with $mr_{s}=5$ and various $mM$.
We find that $\frac{\omega}{m}$ decreases as a function of $mM$.
In Table II, with $mM=1.5$ and various $mr_{s}$, it can be easily seen that
the frequency $\frac{\omega}{m}$ increases with respect to $mr_{s}$.
We also mention that effects of parameters on the frequency
is qualitatively the same as cases
with Dirichlet conditions in the limit $r_{s}\gg M$ \cite{Stationary}.

\renewcommand\arraystretch{2.0}
\begin{table} [h]
\centering
\caption{The frequency $\frac{\omega}{m}$ with $mr_{s}=5$ and various $mM$}
\label{address}
\begin{tabular}{|>{}c|>{}c|>{}c|>{}c|>{}c|>{}c|}
\hline
$~mM~$ &~0.5~& ~1.0~& ~1.5~& ~2.0~& ~2.5\\
\hline
$~\frac{\omega}{m}~$ & ~0.98315836~ & ~0.95300241~& ~0.91654092~& ~0.87691533~& ~0.84537691\\
\hline
\end{tabular}
\end{table}

\renewcommand\arraystretch{2.0}
\begin{table} [h]
\centering
\caption{The frequency $\frac{\omega}{m}$ with $mM=1.5$ and various $mr_{s}$}
\label{address}
\begin{tabular}{|>{}c|>{}c|>{}c|>{}c|>{}c|>{}c|}
\hline
$~mr_{s}~$ & ~4~& ~5~& ~6~& ~7~& ~8\\
\hline
$~\frac{\omega}{m}~$ & ~0.90082899~& ~0.91654092~& ~0.92813679~& ~0.93681321~& ~0.94352675\\
\hline
\end{tabular}
\end{table}

We give a physical discussion about the dependences given in Table I and II.
The scalar field equation (5) can be expressed in the following form
\begin{gather}
\psi''+(\frac{2}{r}+\frac{f'}{f})\psi'-(m^2-\frac{\omega^2}{f})\frac{\psi}{f}=0.\tag{29}
\end{gather}
From equation (29), the effective scalar field mass is defined as
$m^2_{eff}=m^2-\frac{\omega^2}{f}=m^2(1-\frac{\frac{\omega^2}{m^2}}{1-\frac{2M}{r}})$.
It is known that negative scalar field mass
usually makes the scalar field easier to condense \cite{st1}.
We show that the effective scalar field mass
could be negative around the star surface.
A negative effective scalar field mass around the star may trigger the scalar field field to condense.
At the infinity, the effective scalar field mass asymptotically approaches
a nonnegative value $m^2-\omega^2\geqslant 0$, which could serve as a potential to confine scalar field hairs \cite{Sanchis}.
At the star surface $r=r_{s}$, the effective scalar field mass
is $\tilde{m}^2_{eff}=m^2(1-\frac{\frac{\omega^2}{m^2}}{1-\frac{2M}{r_{s}}})$.
In horizonless spacetimes, there is $0<1-\frac{2M}{r_{s}}<1$.
For fixed $1-\frac{2M}{r_{s}}$, a large enough $\frac{\omega^2}{m^2}$
leads to a negative $\tilde{m}^2_{eff}$.
With larger mM and fixed $mr_{s}$, $f(r_{s})=1-\frac{2M}{r_{s}}=1-\frac{2mM}{mr_{s}}$
decreases and we need a smaller $\frac{\omega^2}{m^2}$ to get the negative $\tilde{m}^2_{eff}$.
In the case of larger $mr_{s}$ and fixed mM,
$f(r_{s})=1-\frac{2M}{r_{s}}=1-\frac{2mM}{mr_{s}}$
increases and we need a larger $\frac{\omega^2}{m^2}$ to get the negative $\tilde{m}^2_{eff}$.
So it is natural that the frequency $\frac{\omega}{m}$ may decrease with increasing mM and
increase with increasing $mr_s$,
as is shown by numerical data in Table I and II.

\section{Conclusions}

We studied the gravity model composed of stationary scalar fields
and horizonless neutral compact stars in the asymptotically flat background.
We imposed Neumann surface boundary conditions for scalar fields.
We analytically obtained bounds on the frequency of stationary scalar field
as $max \{1-\frac{2M}{r_{s}},1-4m^2M^2\} \leqslant\frac{\omega^2}{m^2}\leqslant 1$,
were $\omega$ is the scalar field frequency, m is the scalar field mass,
M cooresponds to the star mass and $r_{s}$ represents star radii.
Below the bottom bound or above the upper bound, stationary scalar hairs cannot form outside
horizonless Neumann compact stars.
For certain discrete frequency between the bounds, we numerically obtained
solutions of stationary scalar hairy Neumann compact stars.
We also investigated effects of model parameters on the discrete scalar field frequency.

\begin{acknowledgments}

This work was supported by the Shandong Provincial Natural Science Foundation of China under Grant
No. ZR2018QA008. This work was also supported by a grant from Qufu Normal University of China under Grant
No. xkjjc201906.

\end{acknowledgments}

\end{document}